\documentclass[conference]{IEEEtran}
\IEEEoverridecommandlockouts
\usepackage{cite}
\usepackage{amsmath}
\usepackage{flexisym}
\usepackage{breqn}
\usepackage{amssymb,amsfonts}
\usepackage{caption}
\usepackage{supertabular,booktabs}
\usepackage{algorithmic}
\usepackage{graphicx}
\usepackage{booktabs}
\usepackage{subcaption}
\usepackage{textcomp}
\usepackage{placeins}
\usepackage{longtable}
\usepackage{xcolor}
\def\BibTeX{{\rm B\kern-.05em{\sc i\kern-.025em b}\kern-.08em
    T\kern-.1667em\lower.7ex\hbox{E}\kern-.125emX}}
\begin{document}

\title{An Exploration of Optimal Parameters for Efficient Blind Source Separation of EEG Recordings Using AMICA
\thanks{Funding provided by a gift from the Mathworks, by The Swartz Foundation (Oldfield, NY), and by the NIH grants 5R24MH120037, and 5R01NS047293.}
}

\author{\IEEEauthorblockN{Gwenevere Frank}
\IEEEauthorblockA{\textit{Electrical and Computer Engineering} \\
\textit{University of California San Diego}\\
La Jolla, USA \\
jfrank@ucsd.edu}
\and
\IEEEauthorblockN{Seyed Yahya Shirazi}
\IEEEauthorblockA{\textit{Swartz Center for Computational Neuroscience,} \\\textit{Institute for Neural Computation} \\
\textit{University of California San Diego}\\
La Jolla, USA \\
syshirazi@ucsd.edu}
\and
\IEEEauthorblockN{Jason Palmer}
\IEEEauthorblockA{\textit{Statistics} \\
\textit{West Virginia University}\\
Morgantown, USA \\
japalmer29@gmail.com}
\and
\IEEEauthorblockN{Gert Cauwenberghs}
\IEEEauthorblockA{\textit{Institute for Neural Computation} \\
\textit{University of California San Diego}\\
La Jolla, USA \\
gert@ucsd.edu}
\and
\IEEEauthorblockN{Scott Makeig}
\IEEEauthorblockA{\textit{Swartz Center for Computational Neuroscience,} \\\textit{Institute for Neural Computation} \\
\textit{University of California San Diego}\\
La Jolla, USA \\
smakeig@ucsd.edu}
\and
\IEEEauthorblockN{Arnaud Delorme}
\IEEEauthorblockA{\textit{Swartz Center for Computational Neuroscience,} \\\textit{Institute for Neural Computation} \\
\textit{University of California San Diego}\\
La Jolla, USA \\
\textit{Centre de recherche Cerveau et Cognition} \\
\textit{Paul Sabatier University}\\
Toulouse, France \\
arnodelorme@gmail.com}
}


\IEEEoverridecommandlockouts

\IEEEpubid{\makebox[\columnwidth]{978-1-6654-6819-0/22/\$31.00~\copyright2022 IEEE \hfill}
\hspace{\columnsep}\makebox[\columnwidth]{ }}

\maketitle

\IEEEpubidadjcol

\begin{abstract}
EEG continues to find a multitude of uses in both neuroscience research and medical practice, and independent component analysis (ICA) continues to be an important tool for analyzing EEG. A multitude of ICA algorithms for EEG decomposition exist, and in the past, their relative effectiveness has been studied. AMICA is considered the benchmark against which to compare the performance of other ICA algorithms for EEG decomposition. AMICA exposes many parameters to the user to allow for precise control of the decomposition. However, several of the parameters currently tend to be set according to "rules of thumb" shared in the EEG community. Here, 70-channel AMICA decompositions are run on data from a collection of participants while varying certain key parameters. The running time and quality of decompositions are analyzed based on two metrics, Pairwise Mutual Information (PMI) and Mutual Information Reduction (MIR), and derived recommendations for selecting parameter values are presented. 
\end{abstract}


\section{Introduction}
Brain EEG signals recorded from the scalp are thought to be largely generated by emergent,  locally synchronous field potential  activity in cortical sources, each associated with a patch of adjacent, radially oriented cortical pyramidal cells, volume conducted from  cortex to scalp and linearly mixed at the channel electrodes \cite{nunez1974brain, varela2001brainweb}. This is supported by several biological factors:
\begin{enumerate}
    \item Connections between neurons in close proximity are much denser than those between neurons located further apart \cite{stepanyants2009fractions, stettler2002lateral}.
    \item Inhibitory and glial cell networks have no long-range connectivity \cite{stepanyants2009fractions}.
    \item Connections between the thalamus and cortex are primarily radial  \cite{sarnthein2005thalamocortical, dehghani2010magnetoencephalography}.
\end{enumerate}

As a result, the majority of the cortical contributions to EEG signals recorded at the scalp should be generated by coherent field activity emergent within cortical patches. This means that the effective sources of EEG signals recorded at the scalp are primarily the mixture of locally synchronous or near-synchronous activity in such compact cortical patches. Substantial additional contributions to scalp-recorded signals arise from eye movements and neck muscle activities, as well as line and channel noise.

Scalp-channel recorded mixtures of activity should be well suited for separation using Independent Component Analysis (ICA). Over the years, multiple ICA algorithms have been proposed, many explicitly for the task of source separation in EEG recordings \textbf{\cite{makeig1996}}\cite{ablin2018ortho, ablin2018faster, palmer2012amica}. AMICA, or Adaptive Mixture ICA \cite{palmer2012amica}, has become a popular choice of ICA algorithm for processing EEG data. This is due to its unique abilities to (1) model individual component source densities as mixtures of generalized Gaussians  and (2) when asked, separate the data into subsets of data points each well fit by its own ICA model. AMICA has also been previously shown to perform the best compared to other ICA algorithms when judged on a set of empirical metrics designed to assess the quality of the produced decompositions \cite{delorme2012independent}. However, effective use of AMICA requires users to select values for various adjustable parameters. Currently, these parameters are typically selected based on suggested "rule of thumb" defaults. Here we attempt to provide quantitative arguments for the selections of some key AMICA parameters based on two metrics: Pairwise Mutual Information (PMI) among the derived component time courses, and Mutual Information Reduction (MIR) produced by the decomposition.

\section{Background}

\subsection{Decomposition Metrics}
Two metrics are used to assess the quality of ICA decompositions.

\subsubsection{Mutual Information Reduction (MIR)}

Mutual Information Reduction, or MIR, represents the amount of mutual information removed when ICA is applied to data. More precisely, for the case of ICA,  this is the reduction in mutual information caused when $W$ (the unmixing matrix) is applied to $x$ (the data being decomposed), that is, in linearly transforming the channel time series data to to the discovered independent component (IC) basis.

The reduction in mutual information that results from applying an unmixing matrix $W$ to the data $x$ can be estimated relatively easily using only one-dimensional density models, as pointed out by J. Palmer in \cite{delorme2012independent}. MIR, or Mutual Information Reduction, can be defined as follows:
\begin{dmath}
$$ MIR = I(x) - I(y) = [h(x_1) + \dots + h(x_n)] - [h(y_1) + \dots + h(y_n)] - h(x) + log |det W| + h(x) = log |det W| + [h(x_1) + \dots + h(x_n)] - [h(y_1) + \dots + h(y_n) ] $$
\end{dmath}
where $I(x)$ and $I(y)$ are the mutual information of the source and ICA components, respectively.  Intuitively, this value represents how much mutual information is removed from the data by performing ICA (or any Blind Source Separation, BSS, technique). The above formulation for MIR depends only on the log of the determinant of $W$ and the marginal entropies of $x$ and $y$.

It has generally been shown when sample sizes are large, entropy may be estimated via taking the Riemann sum of a constructed histogram with relatively accurate results. Since EEG data does in fact meet the criteria of having large sample sizes, this approach is used here.

The Riemann sum of the estimated continuous density function correctly integrates to one. Allow the number of bins in the histogram $B$ to be some fixed value. Let  $b_{i}(k), k \in 1,..B$ be the histogram of $x_{i}$, where $b_{i}(k)$ is the number of samples in $x_{i}(t)$ in bin $k$ of the histogram. Let $\Delta_{k}$ be the width of bin $k$ and $N$ be the total number of bins. Then $b_i(k)/(N \Delta_{k})$ estimates $p(x_{i})$ and thus the Riemann sum integrates to one as expected.

ICA seeks to minimize mutual independence and thereby maximize MIR. We can expect that the ICA algorithms that produce the most independent source activities for a dataset (and therefore minimize remaining mutual information) will give the highest MIR values, although the produced MIR values themselves may differ widely across datasets.

\subsubsection{Pairwise Mutual Information (PMI)}

Pairwise Mutual information (PMI) offers a simpler metric of the quality of an EEG decomposition. PMI takes all possible pairs of rows or columns in a matrix and computes the mutual information between those pairs. If we define $x$ as a vector with length $N$ and let $x_{i}(t)$ be a time series that is an element of $x$ of length $n$ and define $M$ as a $nxn$ matrix then mathematically we can express PMI as
\begin{dmath}
[M]_{ij} = I(x_{i}; x{j})=h(x_{i})+h(x_{j})-h(x_{i},x_{j}),  i,j \in 1, \dots, n
\end{dmath}
These marginal entropies may be computed in a manner similar to that described for MIR.

\begin{figure*}[t]
\centering
\includegraphics[width=0.9\linewidth]{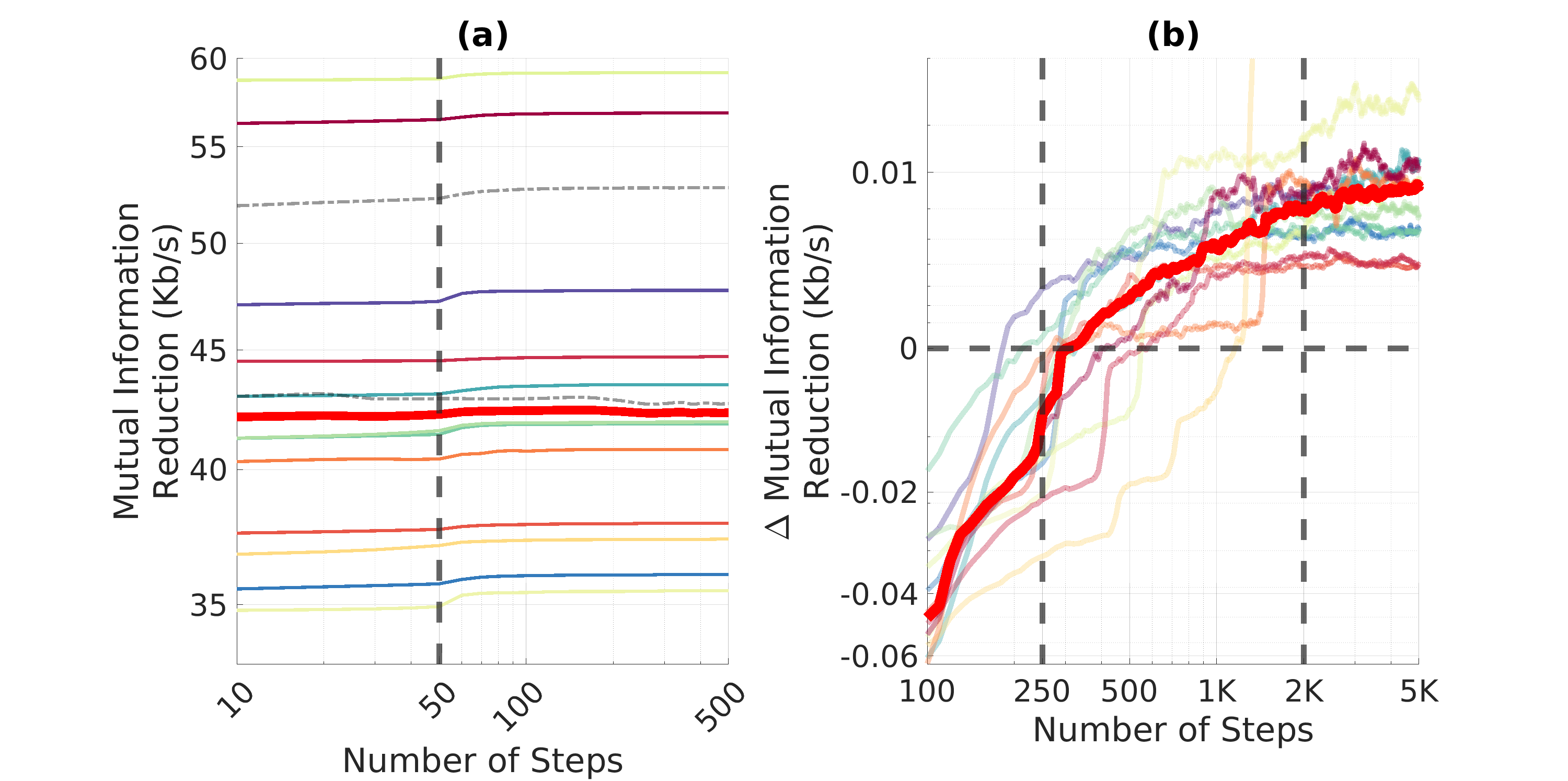}
\caption{(a) MIR (mutual information reduction) for each dataset as a function of the number of iterations. Baseline MIR differs widely across datasets. Dashed vertical line at 50 indicates where newton descent begins. Each color represents a single dataset. The thick red line indicates the median. (b) MIR for each dataset after normalization (see result section). Values are plotted from iteration 100 onward. Each color represents a single dataset.  Dashed vertical line at 2000 indicates the default number of iterations for AMICA, and the dashed line at 250 marks a large change in $\Delta$ MIR (see result section). The thick red line is the median MIR across all datasets for each time point. }
\label{fig:histmirnorm}
\end{figure*}

\begin{figure*}[t]
\centering
\includegraphics[scale=0.3]{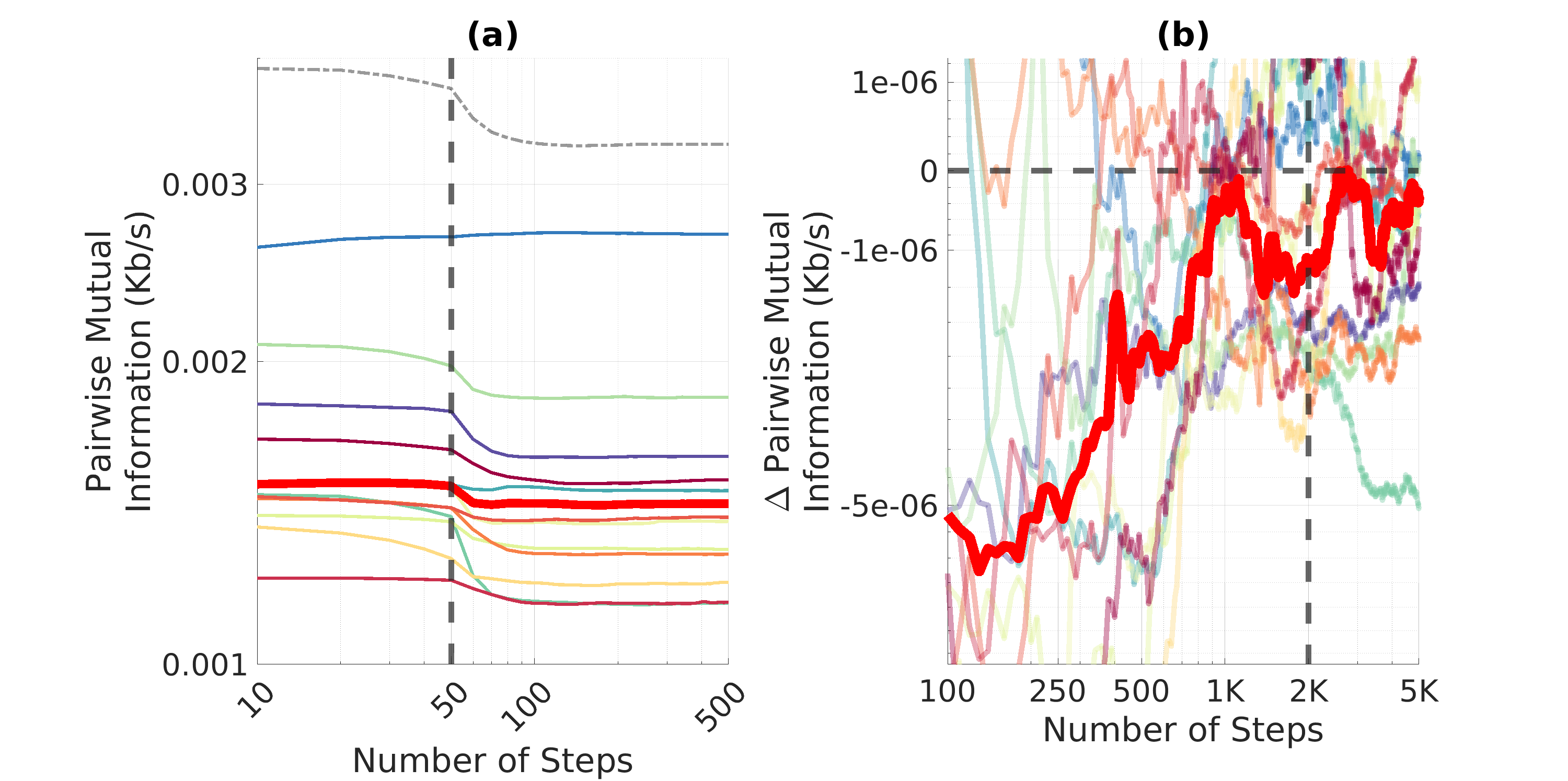}
\caption{(a) PMI (pairwise mutual information) for each dataset as a function of the number of iterations. Baseline PMI differs widely across datasets. The thick red line indicates the median. Dashed vertical line at 50 indicates where newton descent begins.(b) PMI for each dataset after normalization (see result section). Dashed vertical line at 2000 indicates default number of iterations for AMICA. Each color represents a single dataset. Median PMI is shown in the
thick red trace.}
\label{fig:histpminorm}
\end{figure*}

\subsection{The AMICA algorithm}

AMICA is considered to be perhaps the best-performing ICA decomposition algorithm for EEG data decomposition \cite{palmer2012amica, delorme2012independent}. AMICA's performance, quantified by MIR, achieves the highest score among over 20 tested ICA/BSS algorithms \cite{frank2022framework}. AMICA has multiple unique features that set it apart from other ICA algorithms. First, AMICA uses mixtures of Gaussian scale mixture sources to estimate individual source density models, and shapes its learning process based on these. Second, AMICA uses the Amari Newton optimization technique to achieve training with reasonable convergence times for large EEG datasets \cite{ikeshita2022iss2}. AMICA converts second derivative source density quantities to first derivative quantities. Finally, when required AMICA can be used to identify relatively source stationary data subsets and learn ICA models for each \cite{hsu2018}, a capability not tested here. For all analyses, except when otherwise noted, we used the default AMICA parameters (AMICA 1.7, available at https://github.com/sccn/amica). To make analyses comparable across multiple runs, the random seed value in the AMICA parameters was set to the fixed pair of values (123456, 654321). 

\subsection{EEG Data}

We used the same data that we have used in the past to test the performance of a number of ICA/BSS algorithms \cite{delorme2012independent}. Data was collected from more than 20 participants performing a visual working memory task. Participants were asked to first fixate on a small cross presented in the center of a screen for five seconds. Participants were then presented with a sequence of single letters at a rate of about one per second at screen center. Letters were colored either black or green. Black letters were to be memorized; green letters were to be ignored. Participants were then presented with a probe letter, and were tasked with pressing one of two finger buttons indicating whether the probe letter was or was not in the memorized (black letter) set. 400 ms later, participants received auditory feedback as to whether their response was correct. Each subject participated in 100-150 trials.
Data was recorded from 71 scalp channels at 250 Hz/channel after applying an analog 0.1 to 100 Hz pass band filter. All channels were referenced to the right mastoid.  Channels' impedances were kept below 5 KHz. Data was further processed with a custom pipeline implemented in MATLAB using EEGLAB \cite{delorme2004eeglab} as follows: data was high-pass filtered at 0.5 Hz with a linear FIR filter. Data epochs were extracted from 700 ms before to 700 ms after a letter presentation onset. The mean of each epoch was subtracted, and noisy epochs were rejected by visual inspection. Between 1 and 16 epochs were rejected per subject. The MIR and PMI traces for subjects 8 and 10 proved to be erratic, an effect also observed in previous analyses \cite{frank2022framework, delorme2012independent}, and thus, data from these participants were excluded from further analysis. For this analysis, 14 datasets were selected such that half of them would return relatively "good quality" ICA decompositions (as per visual inspection of the IC scalp topographies), and half of them were selected as exhibiting relatively poor ICA decompositions \cite{delorme2012independent}. Each participant dataset comprised data for between 269,000 and 315,000  time samples (~18-21 minutes of data). All data collected from human participants followed an experimental protocol approved by an Institutional Review Board of the University of California San Diego.

\section{Results}

\subsection{Iteration Number}
In practice, AMICA uses a maximum iterations value as its stopping rule. The default maximum used in the EEGLAB implementation of AMICA used here is 2,000 iterations. AMICA is a particularly computationally intensive ICA algorithm, so it would be of benefit if this default stopping parameter could be minimized.

We ran single-model AMICA decompositions for 5,000 iterations on each of the 14 EEG datasets. The seed pair for initialization was the same across all  decompositions. Each ten iterations, the current ICA data unmixing matrix was saved and  MIR and PMI measures were computed on the then-current component activations.

Figure \ref{fig:histmirnorm}a shows that large, stable differences exist between datasets. A grey dashed trace shows the trend for one of the two datasets omitted from further processing and the other omitted dataset's values are greater than the Y-axis scale of this plot. For many datasets, a steeper MIR increase begins at 50 iterations, the (here, default) value at which AMICA newton descent training begins. MIR increases appear to stop after about 200 iterations. In figure \ref{fig:histmirnorm}b, MIR traces are shifted vertically to have zero mean. Here, decompositions beginning with iteration 100 are shown. MIR values on the Y axis are plotted using an inverse log scale to detail how they further evolve during longer decompositions.  The broad red trace shows the median MIR across the selected 12 datasets. (MIR trace colors for individual datasets are identical in both panels). Further MIR increases become smaller after about 250 iterations, and even smaller after about 1000 iterations. Vertical dashed lines indicate the 250 iteration point and the AMICA default stopping iteration (2000). Note the large difference  (approximately 500:1) in y-axis scales in the two panels.

Figure \ref{fig:histpminorm}a plots PMI for each dataset as a function of the number of iterations, and shows that, like for MIR, PMI differs between datasets. Again, the decrease in PMI accelerates when newton descent begins at iteration 50. In figure \ref{fig:histpminorm}b, the traces are normalized and plotted on an inverse log scale in the same way as figure \ref{fig:histmirnorm}b. Values from iteration 100 and onward are plotted. Here, the red line is again the median PMI across all datasets for each time point. The median trace for PMI is somewhat more erratic than MIR. It can be seen that diminishing decreases in PMI occur around 1,000 iterations, again with a slight increase after that point.

\begin{figure*}
   \begin{subfigure}[t]{0.45\textwidth}
    \centering
    \caption{}
    \includegraphics[width=0.9\linewidth]{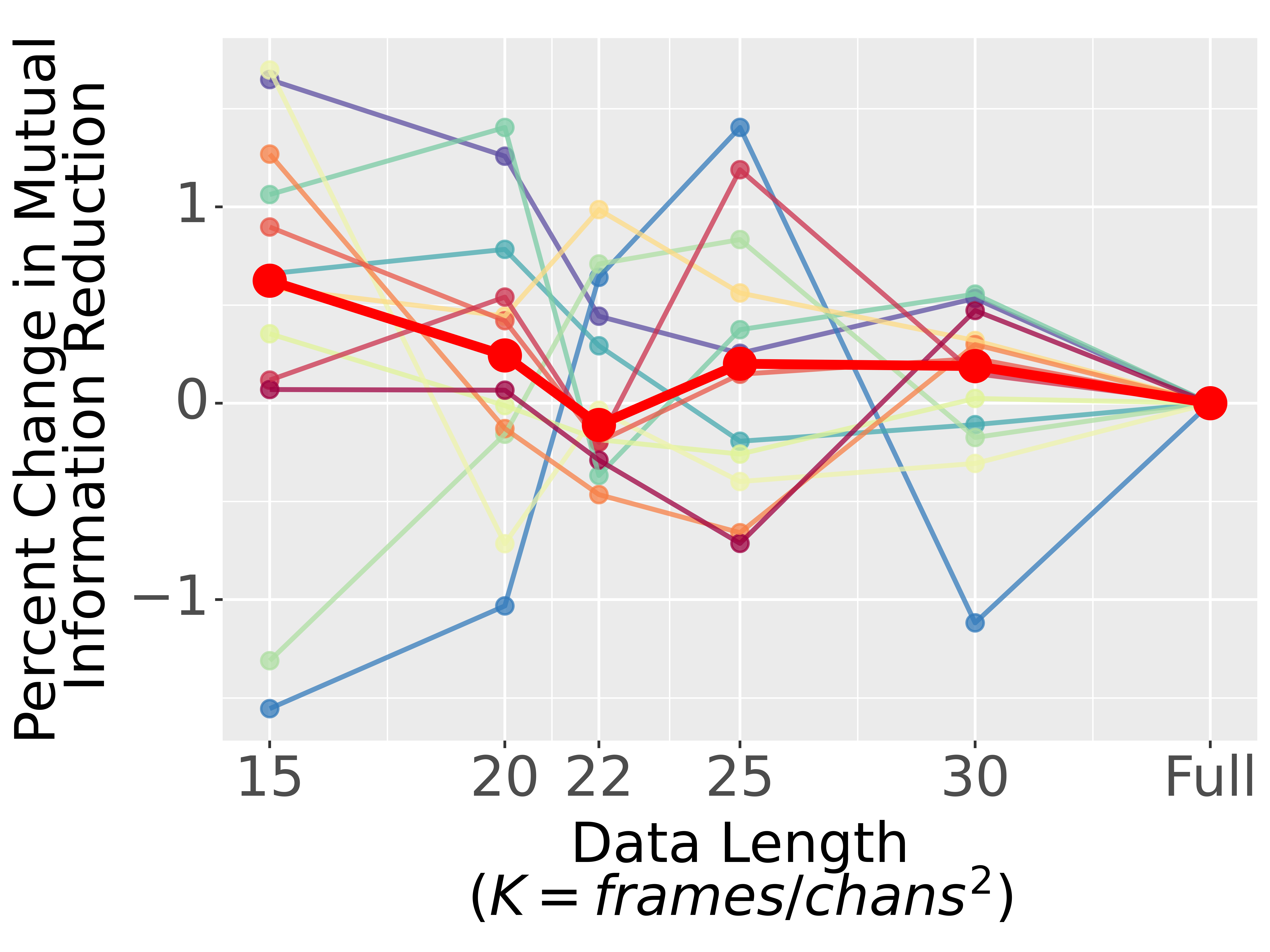}
    \label{fig:mirData}
    \end{subfigure}
    \hfill
    \begin{subfigure}[t]{0.45\textwidth}
    \centering
    \caption{}
    \includegraphics[width=0.9\linewidth]{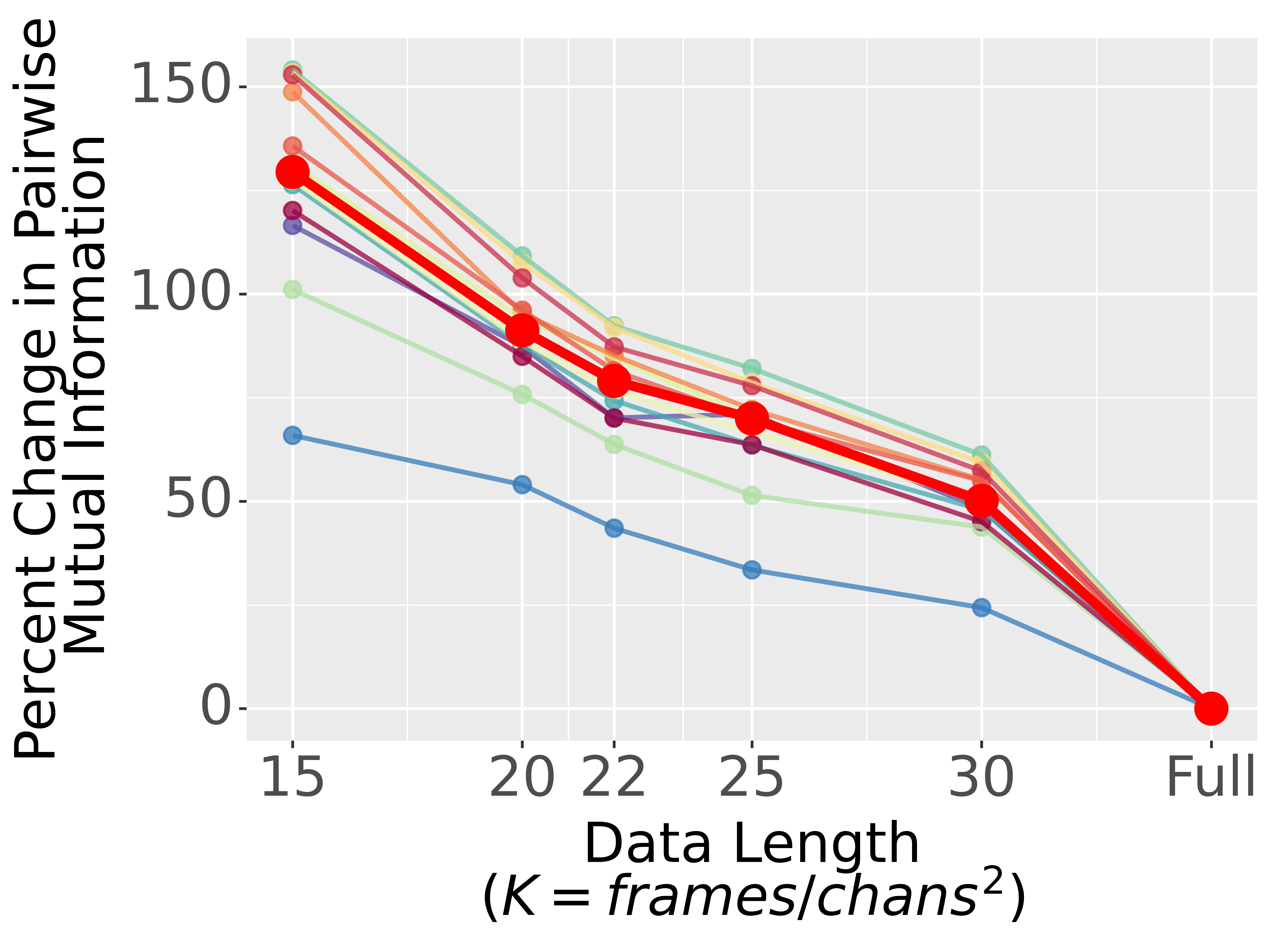}
    \label{fig:pmiData}
    \end{subfigure}
    \vspace{1cm}
    \centering
    \begin{subfigure}[t]{0.45\textwidth}
    \centering
    \caption{}
    \includegraphics[width=0.9\linewidth]{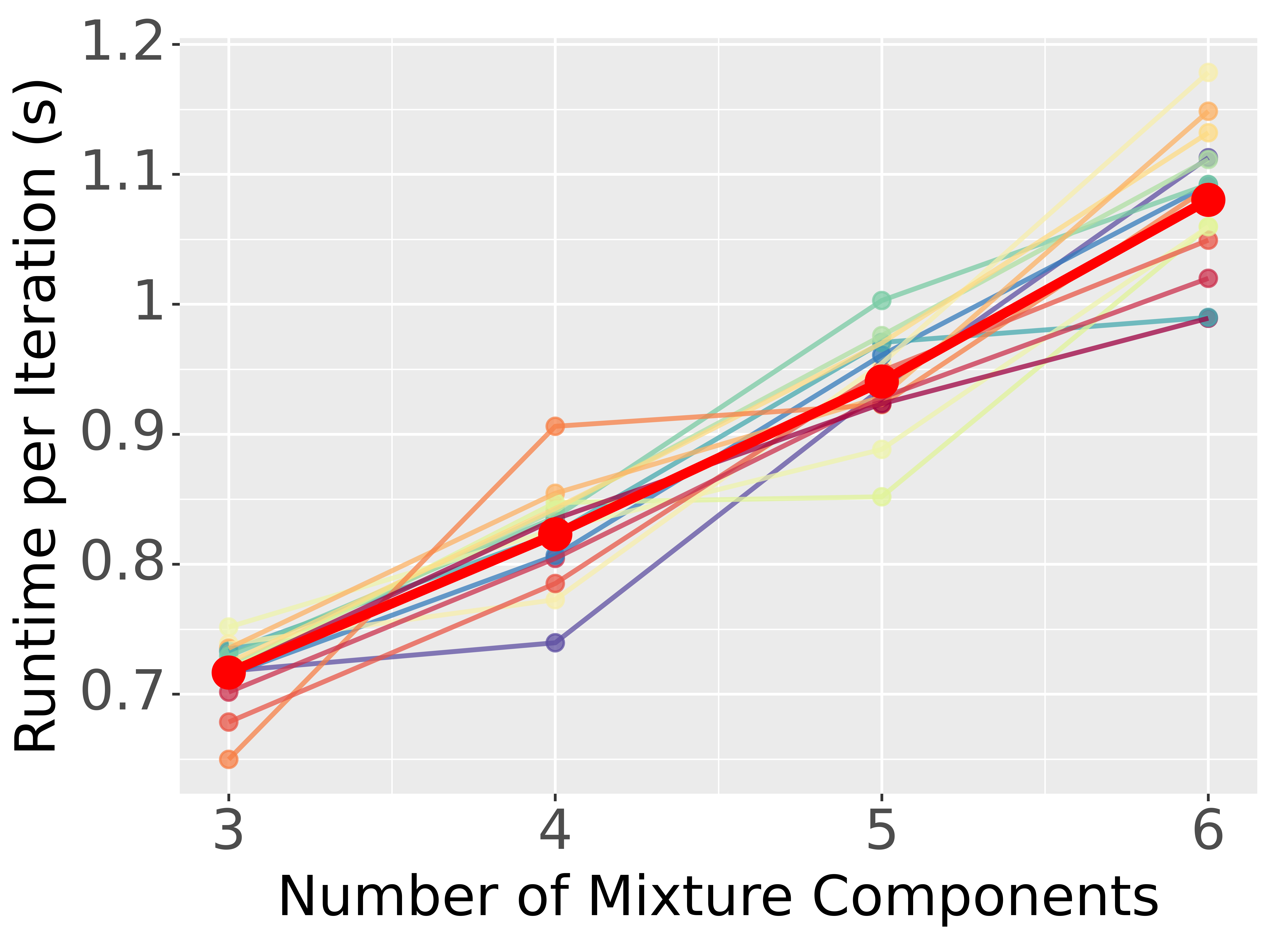}
    \label{fig:nmcRuntime}
    \end{subfigure}
    \hfill
    \begin{subfigure}[t]{0.45\textwidth}
    \centering
    \caption{}
    \includegraphics[width=0.9\linewidth]{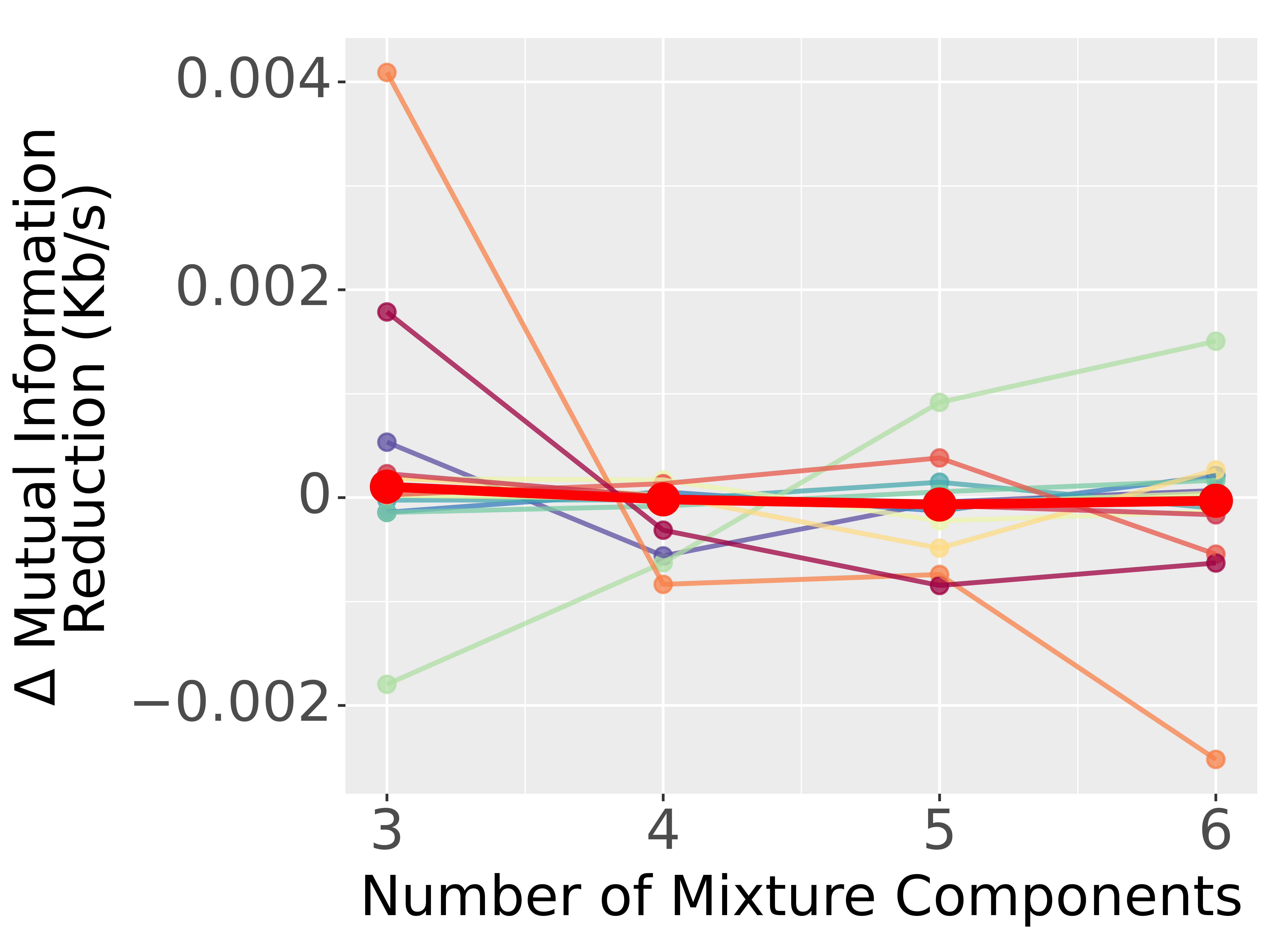}
    \label{fig:nmcMir}
    \end{subfigure}

    \caption{Effect of data quantity and number of mixture models on decompositions. (\subref{fig:mirData}) shows effect of data quantity on MIR (mutual information reduction). (\subref{fig:pmiData}) shows effect of data quantity on PMI (pairwise mutual information). (\subref{fig:nmcRuntime}) shows median run time per iteration across number of mixture models. (\subref{fig:nmcMir}) shows median MIR (mutual information reduction) across number of mixture models.}
\end{figure*}

\subsection{Training Data Quantity}
We have recommended that data decomposed by AMICA has a length of at least $K$ = 22, where $K = \frac{number\ of\ data\ frames}{(number\ of\ channels)^2}$. As the ICA weight matrix contains $(number\ of\ channels)^2$, this is equivalent to suggesting that the training data have at least 22 frames per value in the ICA weight matrix to be learned from the data. To test this heuristic, data epochs were randomly dropped to create datasets with lengths corresponding to several $K$ values, and 5000-iteration AMICA decompositions were performed of these reduced datasets as well as of the full datasets. Epochs in the datasets are composed of 350 frames, thus if the required number of frames to achieve a given value of $K$ was not evenly divisible by 350 the included number of frames was rounded up to the nearest multiple of 350. The mean $K$ value for full datasets was 61.3. The thick red traces in figure 3ab show median MIR and PMI across datasets, as well as  MIR and PMI values for each dataset. Results in figure 3ab are plotted as percent relative to the results for the full data. Figure \ref{fig:mirData} shows changes in MIR have no consistent trend as more data is used. However, figure \ref{fig:pmiData} shows that PMI is decreased as the amount of data is increased.

\subsection{Number of Mixture Models}
A primary source of strength for AMICA compared to other ICA algorithms is its ability to use mixtures of extended Gaussians to model the source probability density distribution of each source. The number of extended Gaussian distributions used in these models is controllable by the ``\texttt{num\_mix\_comp}" parameter. The higher the number, the better an ICA decomposition can model the probability density functions (pdf's) of the EEG sources, increasing the specificity of further adaptations of the source mixing matrix. The number of extended Gaussians may affect both run time and decomposition quality. To test the possible effects of this number, we ran AMICA with 3, 4, 5, and 6 extended Gaussian models, again performing 5,000 iterations with the same seed as in earlier decompositions. Figure \ref{fig:nmcRuntime} shows the median run time per AMICA iteration for each subject (thin traces) and  the median across subjects in red (thick red trace). Run time increases approximately linearly with the number of extended Gaussian models (e.g., by about 50\%  from 3 to 6 distributions). 
Figure \ref{fig:nmcMir} shows mutual information reduction for each participant, normalized to be zero mean, and the median across subjects in red. MIR for individual subjects' fluctuation across mixture model numbers appears to be inconsistent, and no noticeable trend in the change in the median is apparent. The lack of change in MIR doesn’t suggest a difference in near dipolarity of the decomposition created with different numbers of mixture models, but this is still open to further investigation.

\subsection{PMI vs MIR}
\begin{figure}[h]
\centering
\includegraphics[width=0.9\linewidth]{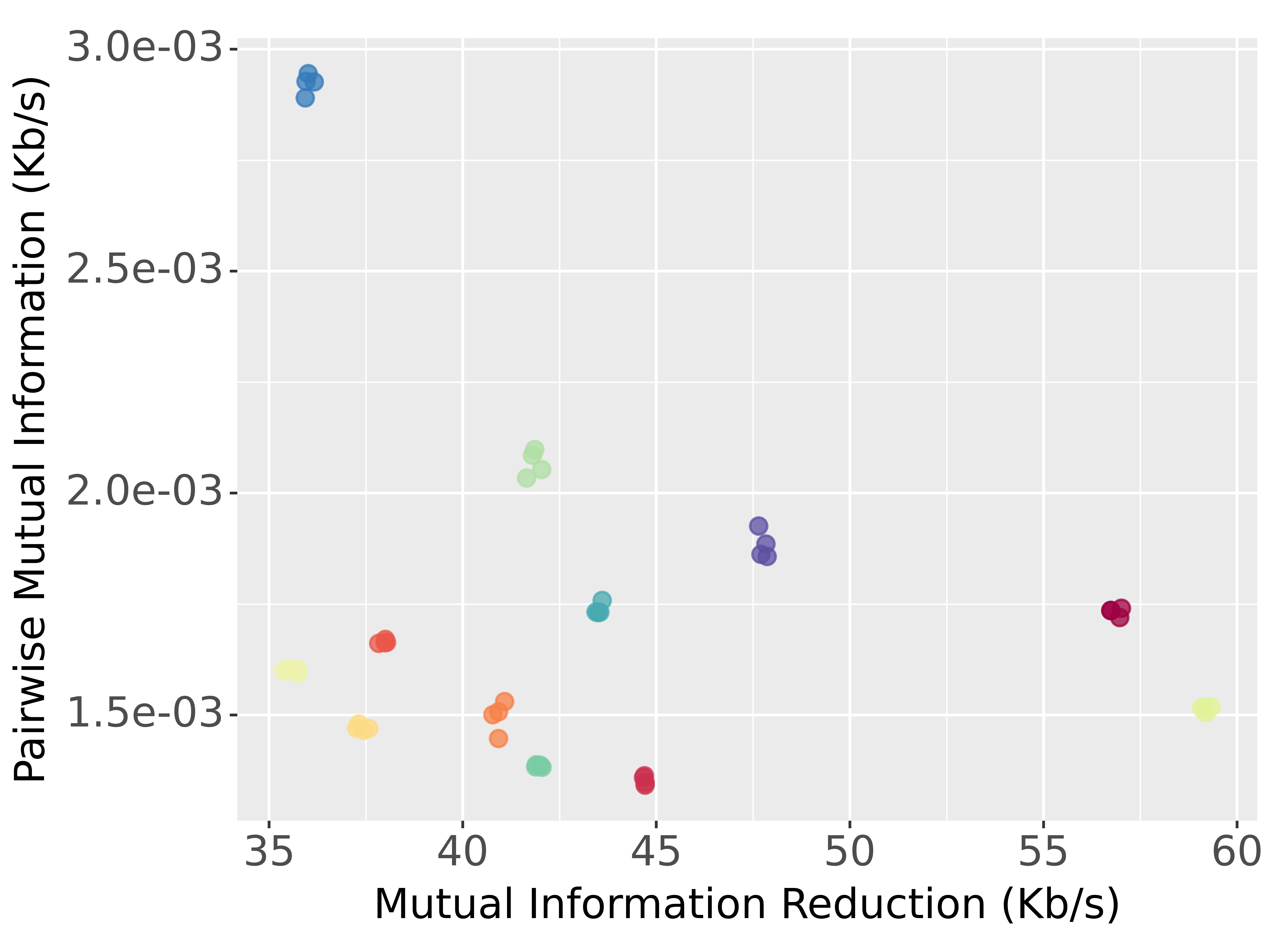}
\caption{PMI (pairwise mutual information) and MIR (mutual information reduction) for 4 randomly selected subsets comprising 75\% of each dataset.}
\label{fig:versus}
\end{figure}

Since both MIR and PMI are measures of association between components, we might expect them to covary. PMI only considers bivariate independence, while MIR takes into account multivariate independence. For each dataset, 75\% of epochs were randomly selected, and decomposition was performed with AMICA for four different trials. 5,000 iterations were run using the same starting seed. Figure \ref{fig:versus} plots PMI versus MIR for each dataset. For each subject, the results of all four trials are shown to  be clustered together. Although the resulting quality of AMICA decomposition is sensitive to the amount of data used, the exact subset of data used in these decompositions appear to be associated with relatively small differences in outcome.

\subsection{Seed Values}
By default, in AMICA the starting weight matrix is randomly initialized. For  numerically complex algorithms, choice of initialization may have important consequences on results -- this is the case for example in deep learning \cite{picard2021torch}. To investigate the effect this initialization has on AMICA decomposition quality, data for a single participant was decomposed using a random seed and 4 fixed seeds for the generation of initial conditions. 5,000 iterations were run, and a single ICA model was fit. For each seed, a decomposition was run five times. MIR values spanned a range of $~0.000511$ Kb/s with a standard deviation of $~0.000148$ Kb/s. PMI values spanned a range of $~8.603883*10^{-7}$ Kb/s and a standard deviation of $~2.087273*10^{-7}$ Kb/s. Thus, compared to the variation observed in Figures \ref{fig:histmirnorm} and \ref{fig:histpminorm}, the change in MIR or PMI based on the seed used is very minor.

\section{Discussion}
We performed a quantitative assay of the effect of iteration number, number of mixture models, and data quantity on the quality of ICA decompositions of 71-channel EEG data produced by AMICA. We found that the most significant changes in resulting MIR and PMI occurred in the first 1000 training iterations. After the initial 1,000 iterations, MIR appeared to fluctuate relatively little, although quite small increases did continue to at least 4,000 iterations. PMI showed a similar trend, with a large decrease in the first 1,000 value and then continued improvement up to at least 3,000 iterations. As for the quantity of EEG data suitable for AMICA decomposition, we verified that for these data a value of $K$ (see results) equal to or greater than 30 was ideal for decomposing EEG data. Increasing the amount of training data was shown to decrease PMI, but had no discernible effect on MIR. Increasing the number of extended Gaussian scale mixtures used by AMICA to model the component pdfs increased run time but did not seem to effect MIR. 
\section{Conclusion}
The analysis included in this paper should provide useful guidance for EEG researchers looking for a better understanding of the effects certain AMICA parameters have on the quality of the resulting decompositions. Future explorations involving a larger array of datasets and parameters are likely to yield further insight. In this work, we did not explore the physiological localizability of the derived cortical component scalp maps. Future work should also explore the data length required to learn more than one AMICA model, each accounting for a learned subset of the training data.

\FloatBarrier

\bibliographystyle{IEEEtran}
\bibliography{IEEEabrv,refs}

\end{document}